\documentclass{aastex}
\usepackage{emulateapj5}
\input{epsf.sty}

\def\solar{\ifmmode_{\mathord\odot}\else$_{\mathord\odot}$\fi} 
\def\kms{km\thinspace s$^{-1}$}   
\def\deg{\ifmmode^\circ\else$^\circ$\fi} 
\def\arcs{\ifmmode {'' }\else $'' $\fi} 
\def\arcm{\ifmmode {' }\else $' $\fi}  
\def\mper{\ifmmode \buildrel m\over . \else $\buildrel m\over .$\fi} 
\def\hper{\ifmmode \rlap.^{h}\else $\rlap{.}^h $\fi} 
\def\sper{\ifmmode \rlap.^{s}\else $\rlap{.}^s $\fi} 
\def\arcsper{\ifmmode \rlap.{'' }\else $\rlap{.}'' $\fi} 
\def\arcmper{\ifmmode \rlap.{' }\else $\rlap{.}' $\fi} 
\def\degper{\ifmmode \rlap.{$^\circ$}\else $\rlap{.}^\circ $\fi} 

\shortauthors{Rosenberg and Schneider}
\shorttitle{Determining the H~I Mass Function}

\begin{document}

\title{The Arecibo Dual-Beam Survey: The H~I Mass Function of Galaxies}

\author{Jessica L. Rosenberg\altaffilmark{1} \& Stephen E. Schneider}
\affil{Department of Astronomy, University of Massachusetts, Amherst, MA 01003}
\altaffiltext{1}{Present address: Center for Astrophysics \& Space Sciences,
Department of Astrophysical and Planetary Sciences, University of Colorado,
Boulder, CO 80309}
\email{jrosenbe@casa.colorado.edu}
\email{schneider@messier.astro.umass.edu}

\begin{abstract}

We use the H~I-selected galaxy sample from the Arecibo Dual-Beam Survey
(Rosenberg \& Schneider 2000) to determine the shape of the H~I mass
function of galaxies in the local universe using both the step-wise maximum
likelihood and the $1/{\cal V}_{tot}$ methods. Our survey region spanned
all 24 hours of right ascension at selected declinations between 8$^\circ$
and 29$^\circ$ covering $\sim$430 deg$^2$ of sky in the main beam. The
survey is not as deep as some previous Arecibo surveys, but it has a larger
total search volume and samples a much larger area of the sky. We conducted
extensive tests on all aspects of the galaxy detection process, allowing us
to empirically correct for our sensitivity limits, unlike the previous
surveys. The mass function for the entire sample is quite steep, with a
power-law slope of $\alpha \approx -1.5$. We find indications that the slope
of the H~I mass function is flatter near the Virgo cluster, suggesting
that evolutionary effects in high density environments may alter the shape
of the H~I mass function. These evolutionary effects may help to explain
differences in the H~I mass function derived by different groups. We are
sensitive to the most massive sources ($log$ M $> 5 \times 10^{10}$
M\solar) over most of the declination range, $\sim1$ sr, and do not detect
any massive low surface brightness galaxies. These statistics restrict the
population of Malin 1-like galaxies to $<5.5\times10^{-6}$ Mpc$^{-3}$.

\end{abstract}
\keywords{galaxies: mass function --- radio lines: galaxies}

\section{Introduction}

One of the main motivations for the Arecibo Dual-Beam Survey (ADBS,
Rosenberg \& Schneider 2000, hereafter Paper 1) was to determine the shape
of the H~I mass function and, in particular, to determine the amount of
mass tied up in low H~I-mass galaxies. Our data indicate that the H~I mass
function is quite steep down to our effective sensitivity limit of about
$3\times10^7 M_\odot$. As parameterized by a Schechter function, we find a
power law slope of $\alpha=-1.5$.

The faint end slope of the H~I mass function has been the focus of
considerable controversy, resulting in uncertainty about the fraction of
the overall hydrogen budget contributed by low-mass galaxies. We suggest
that the differences found between different groups are at least partially
caused by environmental influences, with ram-pressure stripping and merging
in high-density environments resulting in the depletion of low mass H~I
sources.

Most of the shallower slopes ($\alpha\approx-1.2$) for the H~I mass
function have been derived from optically-selected samples in the field
(Huchtmeier 2000, Briggs \& Rao 1993), or in clusters (Briggs \& Rao 1993)
or H~I-selected samples in high density regions like the Canes Venatici
group (Kraan-Korteweg et al.\ 1999), Centaurus A (Banks et al.\ 1999), or
Ursa Major (Verheijen 2000). The results for the faint end slope in the
field have been more varied. Some H~I-selected samples (Zwaan et al.\ 1997) 
have also suggested a slope of $\alpha\approx-1.2$, while other studies 
indicate that the slope might be steeper. Early Parkes survey results 
suggest a slope of $\alpha \approx -1.5$ (Henning et al. 2000; Kilborn et al. 
2001), and our analysis of two earlier Arecibo surveys
(Schneider, Spitzak, \& Rosenberg 1998, hereafter SSR) suggested a steep
faint-end rise similar to that of some optical field galaxy samples
(Loveday 1997; Driver \& Phillipps 1996).

In addition to the effects of environment, we believe a significant source
of discrepancies between surveys arise from differences in the analysis
methodology and the determination of sensitivity limits. One of the most
important innovations of the ADBS is the introduction of ``synthetic"
sources that were carried through the entire data reduction stream in order
to accurately characterize our recovery rate (Paper 1). Most previous
surveys have relied on blanket claims of $N$-$\sigma$ sensitivity without
demonstrating their completeness at the quoted level. In fact, most samples
we have examined fail ${\cal V}/{\cal V}_{max}$ completeness tests (SSR;
Schneider \& Schombert 2000).

The data for this analysis are derived from the Arecibo Dual-Beam Survey
(see Paper 1 for survey details). This is a ``blind" H~I driftscan survey
that covers $\sim$ 430 deg$^2$ in the main beam. The final galaxy tally is
265 sources in the velocity range $-$654 to 7977 \kms. These 265 galaxies
represent all of the sources detected in by-eye examinations of the Arecibo
data that were reconfirmed at Arecibo and/or the VLA. Seven of the galaxies
have M$_{HI} < 10^8$ M$_\odot$, almost as many low mass sources as found in
all earlier blind surveys combined (we use H$_0$ = 75 \kms\ Mpc$^{-3}$
throughout). The average $rms$ noise of the survey spectra is 3.5 mJy so a
$10^8$ M$_\odot$ source with a 50 \kms\ velocity width would be a
5-$\sigma$ detection at 22 $h_{75}^{-1}$ Mpc (1650 \kms).

In \S 2 we describe the survey sensitivity and the relationship between a
galaxy's HI mass and its observed flux as a function of position,
frequency, and distance. In \S 3 we describe the probability of detecting a
galaxy as a function of the profile linewidth and the observed flux. This
section makes use of the ``synthetic" sources inserted in the data to
determine these probabilities. In \S 4 we derive the field mass function
for the ADBS using two standard techniques: the stepwise maximum likelihood
(SWML) and the $1/{\cal V}_{tot}$ methods. We also examine whether the mass
function might be significantly altered by different assumptions like
varying our minimum velocity cutoff, our distance determination method,
or the shape of the completeness function. In \S 5 we discuss the influence
of galaxy density on the shape of the H~I mass function, and show how the
mass function might be affected by making corrections for large scale
structure (in the $1/{\cal V}_{tot}$ method). In \S 6 we examine the limits
that the survey places on the population of galaxies with extremely high
H~I masses, and in \S 7 we summarize our results.

\section{Survey Sensitivity}

The interpretation of H~I surveys requires accounting for variations in
sensitivity as a function of source position and redshift. In addition, all
current blind H~I surveys are only sensitive to low mass sources at low
redshifts, which can potentially introduce large errors in nearby source
distances. Optical surveys have similar issues---caused by vignetting,
K-corrections, and magnitude limits---but these are generally much less
important.

Whenever possible, we have established empirical relationships to describe
our survey's sensitivity and/or applied alternate methods to test the
robustness of our results. In the following sections we describe the noise
levels for our spectra throughout the survey (\S 2.1), and then study the
relationship between a galaxy's HI mass and its observed flux as a function
of position (\S 2.2), frequency (\S 2.3), and distance (\S 2.4).

\subsection{Baseline Noise and Coverage}

The average $rms$ sensitivity for individual spectra, after Hanning
smoothing to a resolution of 32 \kms, was $\sigma_1 = 3.5$ mJy. The
variation around this value was small except for occasional episodes of
heavy broadband interference which occurred in about 2\% of our
observations. Since none of our 265 sources were detected during these
high-noise episodes, we have eliminated them from further consideration.

Over 31\% of the area observed, the ADBS scanned the position only once,
the remainder was covered at least twice. We did not coadd spectra where
they overlapped because there were occasionally slight differences in the
data-taking rate, brief gaps associated with data dumps, and episodes of
broadband interference. All of these small differences would have left us
with a data cube with very complex noise variations if we had coadded the
data. Rather, we examined doubly-covered regions in parallel, using the
duplication of faint sources to help us more reliably identify sources.

Even though the spectra in double-covered regions were not coadded,
examining the data in parallel allowed us to detect fainter sources. This
sensitivity improvement is reflected by our detecting 77\% of our sources
in the 69\% of the survey area that was double-covered. We estimated the
improvement in sensitivity by comparing ${\cal V}/{\cal V}_{max}$
completeness tests (see SSR) for single- and double-covered regions. We
find that double coverage was equivalent to a noise reduction by a factor
of 1.2, giving an effective $rms$ noise for double-covered regions of
$\sigma_2 = 2.9$ mJy.

\subsection{Position Dependence}

Because the ADBS was a driftscan survey, the sensitivity to sources
decreased with increasing declination-offset, $\Delta\delta$, from each
feed's fixed declination. The theoretical ``integrated'' beam of the
survey, which describes the sensitivity decrease, is shown in Paper 1. The
actual sensitivity is more complicated since it depends on the convolution
of each galaxy's H~I distribution with a beam that is not precisely
azimuthally symmetric and which is sampled at intervals that do not
integrate each galaxy's entire flux, $\int S\,dv$, as it passes through the
beam. We determined the galaxies' fluxes ($\int S\,dv$) with follow-up, 
centered
observations at Arecibo and the VLA, and compared these with their
originally observed fluxes ($\int S_{obs}dv$) as a function of their
declination offset. To simplify the notation, we will call these integrated
fluxes or ``signals'' $\cal S$ and ${\cal S}_{obs}$ respectively. We found
that the empirical relationship between the two was well fit by the function:
\begin{equation}
f(\Delta\delta) = {{\cal S}\over{\cal S}_{obs}} \approx 1 + 0.28 \cdot abs(\Delta\delta)^{1.8}
\end{equation}
for $\Delta\delta$ in arcmin. The first sidelobe is at $\Delta\delta \sim
4.4'$ (see Paper 1) where $f$ drops to about 0.2. Beyond $\Delta\delta =
4.4'$ the correction factor is more uncertain, but we detected 20 sources
with offsets up to $\Delta\delta=12'$. The galaxies with large offsets were
all high mass galaxies with extended H~I distributions and had mean
detection fluxes $\sim$10\% of their remeasured values. These 20 sources
are not included in the H~I mass function calculations.

\subsection{Frequency Dependence}

The two line-feed receiver systems used in the ADBS each had variable gain
depending on the redshifted frequency of the H~I signal. We determined the
gain, $g(\nu)$, across the bandpass by examining continuum sources that
fell within the survey region, and (following standard practice at the
observatory) fitted the gain variations with a Gaussian function of
frequency:
\begin{equation}
g(\nu) = 2^{-[2(\nu-\nu _{cen})/52]^2}
\end{equation}
where $\nu _{cen}$ is the center frequency for the feed. For the 21 cm feed
we determined this to be 1408.5 MHz; for the 22 cm feed, 1398.5 MHz.
Both feeds had a half-power frequency response width of 52 MHz.

We determined the statistics for single and double sky coverage by the 21
and 22 cm feeds separately because of the differences in the feed
responses. The 22 cm feed was more often affected by the occasional
broadband interference and had a lower gain at the small redshifts so it
was less sensitive to the lowest-mass sources, which can only be detected
nearby.

\subsection{Distance Estimates}

The ADBS search volume includes the Virgo Cluster. This not only introduces
possible problems with respect to variations in the shape of the mass
function because of the environment, but also causes large uncertainties in
the redshift/distance relationship. We adopt the Tonry et al. (2000) flow
model to derive distances from our measured redshifts, and assumed a Hubble
constant of $H_0=75$ \kms\ Mpc$^{-1}$. We tested the sensitivity of our
results to the choice of flow model by also carrying out the calculations
using a simple distance estimate based on $V_0$, the heliocentric velocity
corrected by $300\cos(b)\sin(l)$ for Local Group motion (de Vaucouleurs et
al. 1977; also see \S 4.3).

The flow correction model reduces distance uncertainties, but within
6$^\circ$ of the center of the cluster large uncertainties are impossible
to avoid so we handle this region separately in our analysis (\S 5.2). We
believe that the Tonry model provides as effective a distance correction as
is available, but it is important to note that all current blind H~I
surveys suffer from detecting low mass sources at small redshifts where the
distance uncertainties are large. It will take a much deeper survey
covering a large volume to improve this situation.

Given the distance of each source we can compute its mass in solar
masses from:
\begin{equation}
 M_{HI} = 2.36\times10^5 D^2 {\cal S},
\end{equation}
(Roberts 1975)
where $D$ is the distance in Mpc. Equivalently, we can predict the expected
noise-free detection signal strength from a source located at any position
in our survey volume as:
\begin{equation}
{\cal S}_{det} = 4.24\times10^{-6}{g(\nu) M_{HI} \over f(\Delta\delta) D^2} .
\end{equation}
Note that since we have an implicit relationship between $D$ and $\nu$, the
detection signal strength can be predicted from the source mass, distance,
and declination offset.

The distinction we make between ${\cal S}_{obs}$ and ${\cal S}_{det}$ is
that the former includes noise and may suffer from peculiarities due to the
specific observational conditions, like asymmetries in the beam shape.
There are several other small effects that we have ignored:

(1) Variations with declination: the sky moves through the telescope beam
by up to 5\% faster or slower at the southernmost and northernmost of our
declination strips. This actually has an even smaller percentage effect on
the maximum observed signal strength because the galaxies dwell within the
beam for several sampling intervals.

(2) Variations with zenith angle: Because of the Arecibo telescope's
design, there are variations with angle from the zenith. Since all of our
observations were made within 11$^\circ$ of the zenith where there was very
little ``spillover,'' these variations are less than a few percent.

(3) Variations with source size: If a source is larger than the beam size,
the measured signal will be smaller than we predict. This proves to be
unimportant for our results, because all sources that ``fill the beam''
have strong signals. The few sources for which there is a more accurate 
flux in the literature have had those values substituted (see Paper 1). 
This is effectively a statement that there are no
galaxies with such low H~I surface densities that long integrations would
be required to detect them, which was shown by the deeper survey of Zwaan
et al.~(1997). Note that this is not necessarily the case for synthesis
observations where the beam size is much smaller, and the H~I surface
brightness sensitivity generally poorer.

The adjustments we have applied are just for the effects that result
in at least tens of percent change.

\section{Sensitivity and Completeness}

The preceding section described how the signal strength and noise levels
vary with location in the survey search volume. We turn next to the
probability of detecting an H~I signal of a particular linewidth $w$ and
integrated flux ${\cal S}_{det}$ in a spectrum with an rms noise
$\sigma$.

Previous surveys have generally attempted to make plausible assumptions
about some signal level to which they believe they should be complete, but
we have found (SSR; Schneider \& Schombert 1999) that such claims do not
pass muster with completeness tests. Since all H~I surveys to date have
relied to varying degrees on human-eye inspection of data, we believe it is
essential to build into the detection step a means of assessing the survey
completeness, ${\cal C}$, empirically.

Our approach to determining the ADBS completeness was to insert a large
number of ``synthetic'' sources throughout the survey data (see Paper 1).
The synthetic profiles were modeled to look very similar to observed H~I
profiles, and were inserted early in the data-processing procedure so that
they would be treated like real sources, suffering, for example, the
effects of automated baselining procedures. The synthetic sources were
given randomized positions, line widths, and line strengths, and their
locations within the data stream were unknown to us during the detection
steps. By determining the rate at which we recover synthetic sources of a
particular line width and signal strength, we can empirically establish the
probability of detecting sources with different linewidths and signal
strengths.

We find that the shape of the completeness function is basically the same
for different line widths $w$, up to some factor in the signal strength
${\cal S}_{det}$. We can write:
\begin{equation}
{\cal C}(w, {\cal S}_{det}, \sigma) =
C\left({\cal S}_{det} \over {\cal N}_{eff}(w)\right)
\end{equation}
where ${\cal N}_{eff}(w)$ is an effective noise level determined
empirically for different line widths. Note that synthetic sources require
no corrections for declination offset or gain corrections so that we know 
the input, noise-free value of the detection flux. In \S 3.1 we 
explain the
derivation of the line width dependence and in \S 3.2 the variations of
completeness with signal strength.

\subsection{Line Width Dependence}

H~I surveys have sensitivity variations depending on the velocity width of
the spectral line. The same total signal $\cal S$ spread over a larger
frequency width, has a lower mean flux density and is harder to
detect.\footnote{Note that there are related problems for optical surveys
that behave in the opposite direction---face-on disk galaxies have a lower
surface brightness and are therefore harder to detect than more edge-on
disks, whereas their H~I line width is narrower, so they are easier to
detect.}

The empirical line-width dependence is slightly different than would be
predicted from basic statistical arguments, but these differences do not
have a strong effect on the final determination of the mass function as we
show in later sections. A statistical model of the noise dependence on line
width ${\cal N}(w)$ is that it grows as the root sum square of the
uncorrelated noise $\sigma$ in individual channels. In this theoretical
case, the noise should grow as
$w^{0.5}$. However, in our analysis of earlier Arecibo surveys (SSR), we
found that the detectability of wide-line sources showed a more-rapid
decline with line width than this noise model predicts. Such a behavior can
be explained as an effect of baselining and other data-processing
procedures that may affect wider line width signals more adversely
than narrow ones.

Based on our empirical results for the synthetic sources, the signal
strength at which the completeness drops to 50\% increases as $w^{0.75}$,
or equivalently we can describe the effective noise  ${\cal N}_{eff}
\propto {\cal N} w^{0.25}$. We arbitrarily set the effective noise value to
match the statistical value at a linewidth of 300 \kms, yielding:
\begin{equation}
{\cal N}_{eff} = {32 \sigma (w_{20}/32)^{0.75}
                      \over {(300/32)^{0.25}}}
\end{equation}
where $\sigma$ is the $rms$ noise in Jy, $w_{20}$ is the line width of the
detected signal in \kms\ measured at 20\% of the peak flux density, and 32
refers to our velocity resolution in \kms. Note that the choice of 300
\kms\ for normalizing ${\cal N}_{eff}$ does not in any way affect our
results since the determination of the completeness $\cal C$ simply uses
this to scale the fluxes for different linewidths.

This dependence on line width is in agreement with what we found for the
surveys of Spitzak \& Schneider (1998) and Zwaan et al. (1997) based on
their ${\cal V}/{\cal V}_{max}$ statistics for different line widths (SSR).
This suggests that the reduced sensitivity to wide-line sources may be
a fairly generic property of H~I surveys.

The effect of the line width dependence is that sources of the same mass
are not necessarily detectable to the same distances. Therefore, in order
to characterize the H~I mass function, we need to detect a sufficiently
large number of sources so that we have a representative sample of the
different characteristics of galaxies in each mass range.

\subsection{The Completeness Function}

We determined the probability of detecting sources---the completeness $\cal
C$---by the rate at which we recovered the synthetic sources as function of
their signal strengths. We initially examined different line width ranges
separately, but the shapes of the curves were similar, so we folded the
data together, scaled by the effective noise, to empirically estimate
${\cal C}({\cal S}_{det}/{\cal N}_{eff})$. Again note that we are studying
the synthetic sources so we know the input, noise-free value of the detection 
flux. The resulting completeness function is shown in Figure 1.

The figure shows that the edge between detectability and non-detectability
is not sharp. This is a combination of at least two effects: (1) noise
added to sources near the limit of detectability of the survey may push
sources above or below the detection threshold; and (2) the interaction of
a particular line shape with the detection methods, whether automated or
conducted by-eye, will introduce some uncertainty in detections near the
nominal ``limit.''  The dashed line in the figure is an error function fit
to the data. An error function is the expected result when Gaussian noise
falls on top of an underlying signal.

\centerline{\epsfxsize=0.9\hsize{\epsfbox{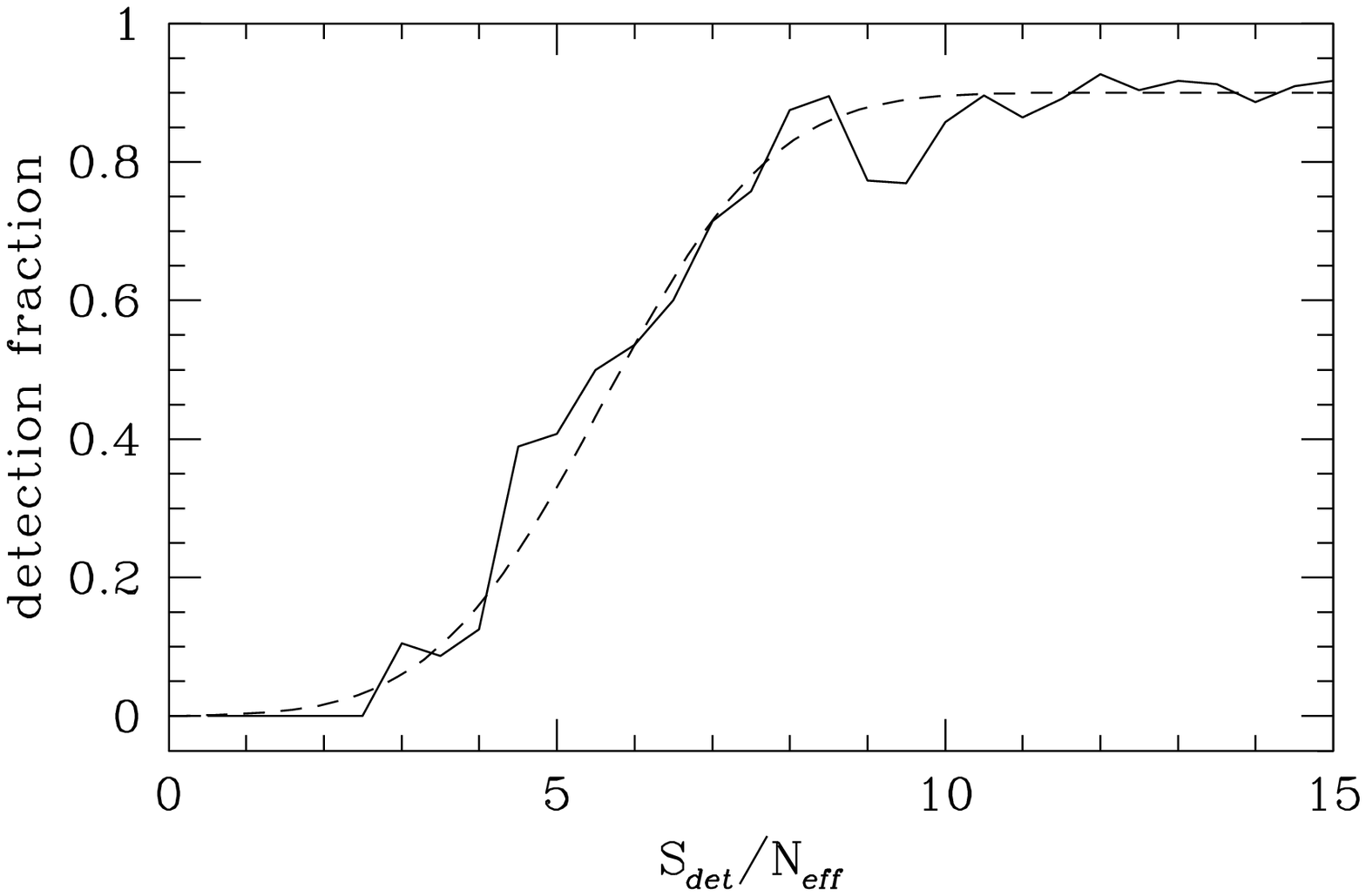}}}

\vspace{0.1in}
\figcaption{The relationship between completeness and 
${\cal S}_{det}/{\cal N}_{eff}$. This function was determined using 
the ``synthetic'' sources from Paper 1. The dashed line is an error 
function fit to the data.}
\vspace{0.2in} 

We found that the detection fraction never reached 1.0, mostly because we
would occasionally lose bright sources near interference spikes or when the
automated baselining procedure did not work well (see Paper 1). We found
that we could not isolate this effect to a few individual frequencies, so
we leave it as an overall correction that will make our estimated source
densities about 10\% higher.

Because of the characteristics of gaussian noise, any cutoff in 
signal-to-noise inevitably includes some sources below the cutoff limit
and excludes some sources above it. One can approximate a step-function
sensitivity by detecting sources to a deep limit and then using high
sensitivity follow-up observations to eliminate all sources below where the
roll-off in completeness becomes significant. Unfortunately, this would
necessitate using a very high cutoff level (${\cal S}/{\cal N}\approx10$),
and ignoring a large amount of data.

Another approach might be to choose a lower cutoff, like 5- or 7-$\sigma$, 
and ignore all sources that prove to be weaker than this limit upon
follow-up observations. This approach will exclude less data than a
high-sigma cutoff, but it has the disadvantage of uncertain effects due to
the incompleteness of low-flux sources.

With knowledge of the completeness function, it is possible to retain
all detected sources when estimating the mass function. We know the
fraction of sources detected at any signal-to-noise level, so we can
correct our estimates of the volume of space searched. Essentially,
where a source's signal strength is so weak that our probability of 
detection is half, we have effectively searched half as much volume
at that distance.

\subsection{Completeness and ${\cal V}/{\cal V}_{max}$}

The ${\cal V}/{\cal V}_{max}$ test (Schmidt 1968) is often used to
determine whether a survey's sensitivity has been properly defined. The
${\cal V}/{\cal V}_{max}$ test assumes a step-function sensitivity cutoff
and uniformly distributed sources. Given these conditions, the average of
the volumes interior to all sources should equal half of the maximum volume
within which these sources could have been detected: $\overline{{\cal
V}/{\cal V}_{max}} = 0.5$.

If we allow for a roll-off in completeness, some sources will be detected
that are actually weaker than whatever nominal sensitivity limit is chosen,
and they will have ${\cal V}/{\cal V}_{max}>1$. It is no longer clear
what value of $\overline{{\cal V}/{\cal V}_{max}}$ should result from
such a sample, or whether the test is applicable.

Actually, it is straightforward to calculate the expected mean value of
${\cal V}/{\cal V}_{max}$ given the completeness function. We carried out a
simple Monte-Carlo calculation for 10,000 uniformly-distributed sources,
randomly selected according to the value of the completeness function for
their predicted fluxes at their assigned distances. Defining ${\cal
V}_{max}$ in terms of the limiting distance for a source with a flux at the
50\% completeness level, we expect $\overline{{\cal V}/{\cal V}_{max}} =
0.61$.

The detected ADBS sources have $\overline{{\cal V}/{\cal V}_{max}} = 0.60$
when measured in the same way, which is in excellent agreement with the
expected value. This agreement shows that the real detected sources behave
in much the same way as predicted by our completeness function, which was
based purely on the synthetic sources.

\section{The H~I Mass Function}

\subsection{Two Methods for Determining the Mass Function}

Given our results for the variation of signal strength within our search
volume, and the probability of detecting sources of a particular signal
strength and linewidth, determining the H~I mass function is, in principle,
a straightforward matter. We estimate the mass function using two
well-known techniques: the ``$1/{\cal V}_{tot}$'' method (see SSR) and the
step-wise maximum likelihood (SWML, Efstathiou et al.\ 1988) method.

These two techniques are complementary in several respects:

(1) The SWML method is formulated to be independent of large scale
structure effects. However, the SWML method assumes that the shape of the
mass function is the same everywhere, which is particularly questionable
for H~I because of gas-stripping and merging in high density regions. The
$1/{\cal V}_{tot}$ method is simpler in concept and makes no prior
assumptions about the uniformity of the mass-function shape.

(2) With the $1/{\cal V}_{tot}$ method, the overall normalization of the
mass function is directly determined. Special techniques are used to
normalize the results from the SWML method, which are not easily adapted to
the complex positional and distance dependencies in an H~I survey. Note
that for neither method does the normalization affect the shape.

(3) The $1/{\cal V}_{tot}$ method requires the detailed knowledge of survey
sensitivity over the entire search volume as worked out in \S 2 and 3 to
calculate the total volume (${\cal V}_{tot}$) in which a source might have
been detected. By contrast, the SWML method only requires that we find the
maximum distance at which a source could have been detected {\it at its
detected position}---this is simpler, since we can just scale the detected
flux for distance and frequency dependencies.

(4) We have further simplified the SWML calculation relative to the
$1/{\cal V}_{tot}$ method by assuming a sharp cutoff in sensitivity at 7
times the effective noise ${\cal N}_{eff}$ predicted for a source of that
linewidth. In part, this provides a test of whether our sensitivity
``roll-off'' might have an important effect on the shape of the mass
function.

The total volume or $1/{\cal V}_{tot}$ method was originally proposed by
Schmidt (1968). ${\cal V}_{tot}$ is the total volume within 
which a source {\it could have been} detected. If we take all the
detected sources within a particular mass range summing up $\sum 1/{\cal
V}_{tot}$ gives us a direct estimate the number density of such sources.

Because we have determined the detailed shape of the completeness function,
we are able to more accurately estimate the total effective search volume
by weighting each position according to the probability of detecting a
source there, as discussed in \S 3.2. The detectable volume for a galaxy is
therefore:
\begin{eqnarray}
{\cal V}_{tot}\!=\!\!\!\!\!\! \sum_{\sigma=\sigma_1,\sigma_2}\!\!\!\!\! A(\sigma) 
\!\!\! \int_{\Delta\delta=-4.4'}^{4.4'}\!\! \int_{D=v_{min}/H_0}^{v_{max}/H_0}
\!\!\!\!\!\!\!\!\!\!\!\!\!\!\!\!\!\!\!\!\!\! {\cal C}(w,{\cal S}_{det},\sigma) D^2\, dD\, d(\Delta\delta)
\end{eqnarray}
where $A(\sigma)$ is the angular extent of the survey with the rms level
$\sigma_1$ or $\sigma_2$, for either single or double coverage, and $\cal
C$ is the completeness function described in \S 3. The distance variable
implicitly incorporates the assumed redshift--distance relationship and the
gain dependencies on frequency. The integration over distance ranges from a
minimum velocity $v_{min}$ below which confusion with Galactic H~I and high
velocity clouds are likely to make source detection difficult to the
maximum redshift covered by our spectra, 7977 km s$^{-1}$. Positional and
frequency dependencies are carried implicitly within ${\cal S}_{det}$ which
varies with distance, declination offset ($\Delta\delta$), and frequency,
as discussed in \S 2. In practice, we carry out this integration by summing
over small intervals and calculating the predicted value of ${\cal
S}_{det}$ at each position and velocity.

The SWML method (Efstathiou et al.~1988) was designed to directly remove
the effects of density variations caused by large-scale structure. It
divides the mass function into a series of bins and solves for the most
likely set of relative weights for the bins. The likelihood function for 
each source
is the ratio of the weight of its own mass bin to the sum of the weights
of all of the mass bins in which a source, with the same redshift and 
limiting flux, could have been detected. By taking
this ratio of weights,
density effects are divided out.  The total likelihood is the
product of the likelihoods for all of the sources. A set of mass-bin
weights that maximizes the total likelihood is found iteratively.

By its design, the SWML method allows each source to have a different flux
limit. The method is usually used in optical surveys where the limiting
flux is uniform, but there is nothing in the design of the method or most
implementations of it that requires this. We use a version of the method
from Kochanek et al.~(2001).

The likelihood function in the SWML method assumes a sharp cutoff in the
sensitivity. It could probably be adapted to include the completeness
function, but as noted in the introduction to this section, we use a fixed
``completeness limit" for the SWML method rather than trying to modify the
routine. We use only the 210 sources brighter than 7 times the effective
noise ${\cal N}_{eff}$. Because the completeness is not 100\% at this
limit, we will tend to slightly underestimate the number of sources near
the flux limit.

Because the H~I source sensitivity varies over our search volume depending
on a variety of parameters (\S 2), it is difficult to normalize the results
from the SWML method. Instead, we scale the results to match the $1/{\cal
V}_{tot}$ results for high-mass ($>10^9 M_\odot$) sources. The high mass
sources were detectable over most of the potential search volume, so the
uncertainties in their density are small. Note again that this does not
affect the calculated shape of the mass function derived by the SWML
method.

We show in the remainder of this section that the two methods display
substantially similar H~I mass functions. Since the two methods use
different approaches, this provides some reassurance that there is not a
fundamental error in one of our approaches. Of course, both methods make
many of the same assumptions about sensitivity, and we attempt to test
those assumptions in various ways in subsequent sections.

For the $1/{\cal V}_{tot}$ calculation of the mass function we include 233 
galaxies, all of the galaxies
confirmed in the Arecibo and VLA follow-up that were originally detected
within 4.4\arcmin\ of the center of the main beam and that are $> 6^\circ$
from the center of Virgo. We recognize that this
is a small sample of galaxies compared to optical surveys, however, we feel
we have the best characterizations of HI survey sensitivity made to date,
so our conclusions are as strong as the data allow.

\subsection{The H~I Mass Function}

Figure 2 shows the H~I mass function excluding the
central $6^\circ$ of Virgo based on the SWML method (filled circles) and
$1/{\cal V}_{tot}$ method (open squares). The curves in the figure are
Schechter (1976) functions:
\begin{equation}
\Phi(M) = {dn\over d\,\log M}=\Phi_\ast \ln 10 (M_{HI}/M_\ast)^{\alpha+1}
e^{-M_{HI}/M_\ast}
\end{equation}
where $\alpha$ is the power-law slope of the faint end, $M_\ast$ is the
characteristic turn-over mass, and $\ln (10/e) \Phi_\ast$ is the density
per mass decade at $M_\ast$. The minimum $\chi^2$ fit to the SWML points
gives: $\alpha=-1.53$, $\log(M_\ast/M_\odot)=9.88$, and $\Phi_0=0.005$
Mpc$^{-3}$. (As noted earlier, the $\Phi_0$ normalization value is
ultimately based on scaling the SWML result to the $1/{\cal V}_{tot}$
results.) The dashed line shows the mass function fit from Zwaan et al.
(1997) from their Arecibo survey: $\alpha=-1.2$,
$\log(M_\ast/M_\odot)=9.80$, and $\Phi_0=0.0059$ Mpc$^{-3}$ (adjusted to
$H_0 = 75\ $\kms\ Mpc$^{-1}$).

\centerline{\epsfxsize=0.9\hsize{\epsfbox{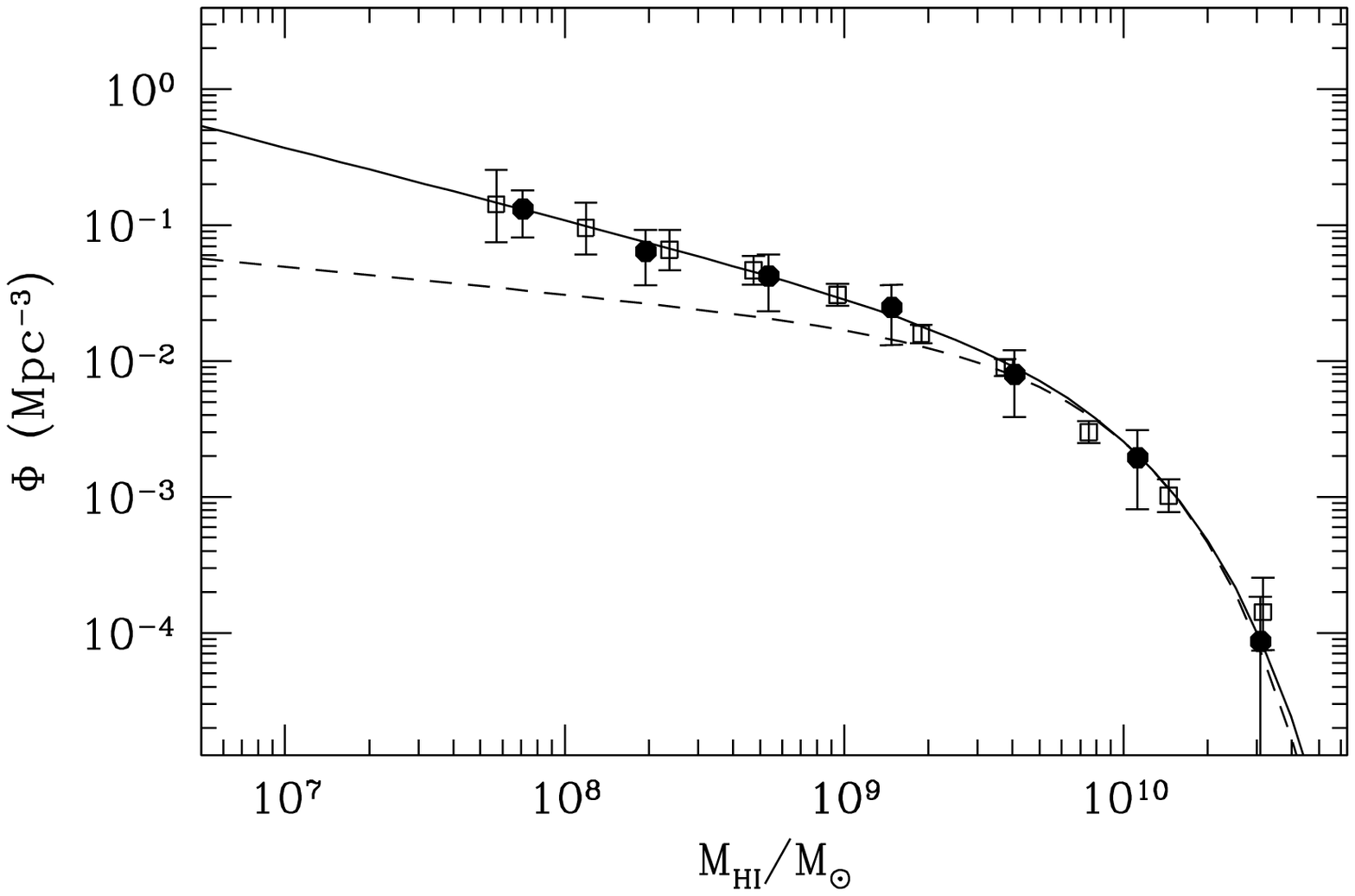}}}

\vspace{0.1in}
\figcaption{The H~I mass function outside of the Virgo cluster core.
The mass function is shown using both the $1/{\cal V}_{tot}$ method
(open squares) and the step-wise maximum likelihood method (solid circles).
The solid line is the best-fit Schechter function: $\alpha=-1.53$,
$\log(M_\ast/M_\odot)=9.88$, and $\Phi_\ast=0.005$ Mpc$^{-3}$. The dashed
line shows the Zwaan et al (1997) Schechter model fit to the AHISS data.
Error bars indicate ``1-$\sigma$'' uncertainties; see text.}
\vspace{0.2in} 

The ``1-$\sigma$'' error bars in Fig. 2 for the $1/{\cal
V}_{tot}$ method are based on small number statistics (Gehrels 1986). The
bins were selected by an automatic procedure that attempted to keep at
least four galaxies in each bin unless the bin widths would otherwise
become very small or large ($\Delta log$(M/M\solar) $> 0.4$), except at the
bright end where a single galaxy was allowed to define a bin. The error
bars for the SWML method include an internal estimate of the uncertainty in
the normalization and are highly correlated by the nature of the method.

It is apparent from the figure that the two approaches yield very
similar-shaped mass functions. It is also evident that our mass function is
significantly steeper than the $\alpha=-1.2$ power law found by Zwaan et
al.~(1997).

\centerline{\epsfxsize=1.0\hsize{\epsfbox{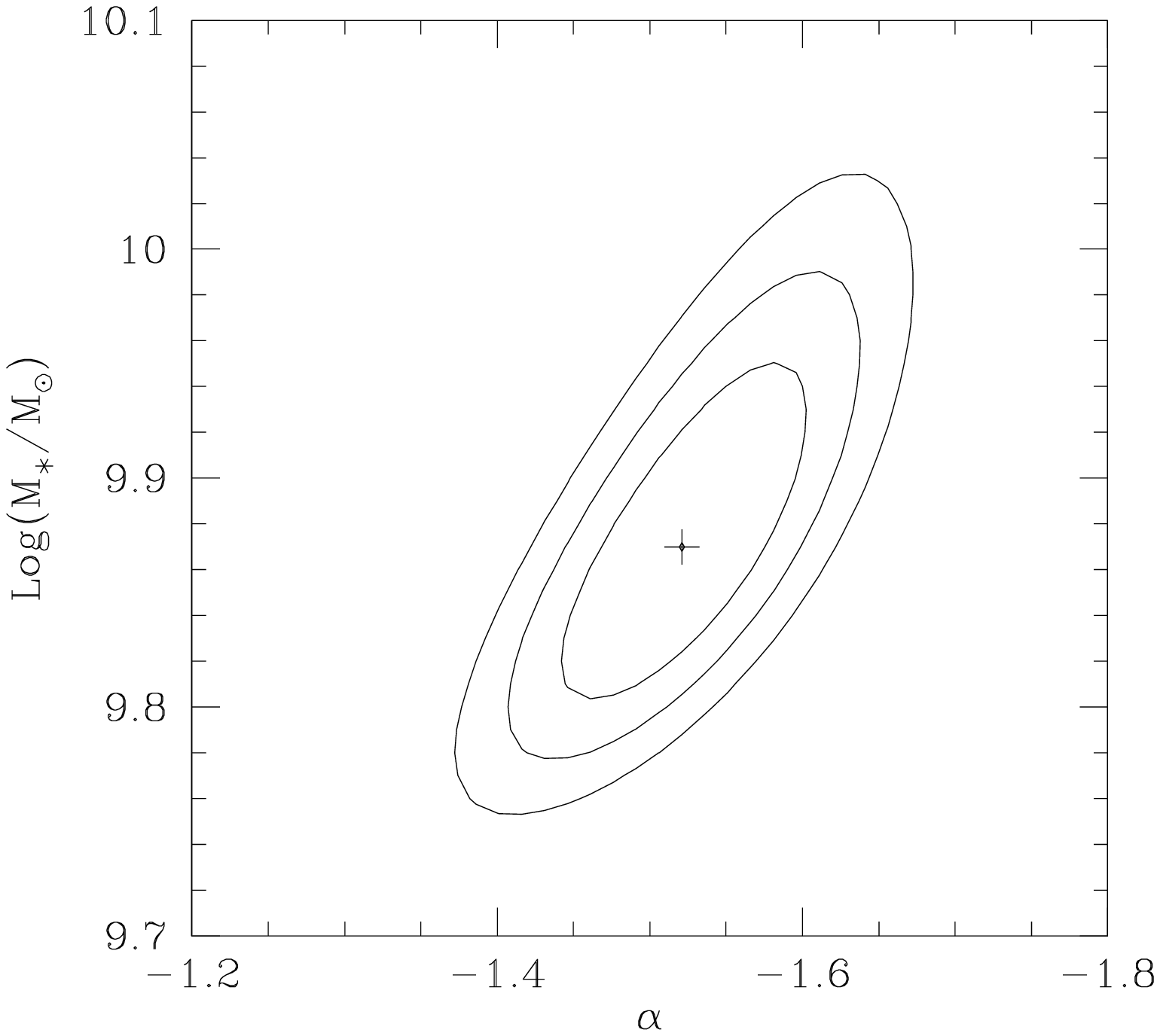}}}

\vspace{0.1in}
\figcaption{The chi-square contours of the Schechter function fit to the
SWML determination of the H~I mass function (shown in Figure
2). The contours represent 1, 2, and 3 $\sigma$.}
\vspace{0.2in} 

The shallow $\alpha=-1.2$ power law is in good agreement with our results
at high masses, but it becomes successively worse at lower masses. Figure
3 shows the probable range of Schechter function parameters
based on a chi-square goodness of fit to the SWML calculated mass function.
Part of our difference from Zwaan et al.~probably comes from our better
number statistics; they had detections of only 66 sources, and only 51 with
$\Delta\delta<4.4'$. Our sample contains 7 sources with H~I masses $< 10^8$
M$_\odot$ as compared with 2 (for H$_0$ = 75 \kms) in Zwaan et al. (1997),
2 in Kilborn et al. (1999), and 4 in Spitzak \& Schneider (1998).

We show in \S4.3 that no significant changes arise when we re-examine our
data set in a variety of ways. Large scale structure within our survey
region also does not appear to bias our results (\S4.4). However, there
{\it are} indications of differences in the faint end slope when a survey
is isolated to a cluster region (\S4.5).

\subsection{Effects of Analysis Procedures}

There are many small choices in determining the H~I mass function that
might affect our results. We find, however, that our result is
robust.

\centerline{\epsfxsize=0.9\hsize{\epsfbox{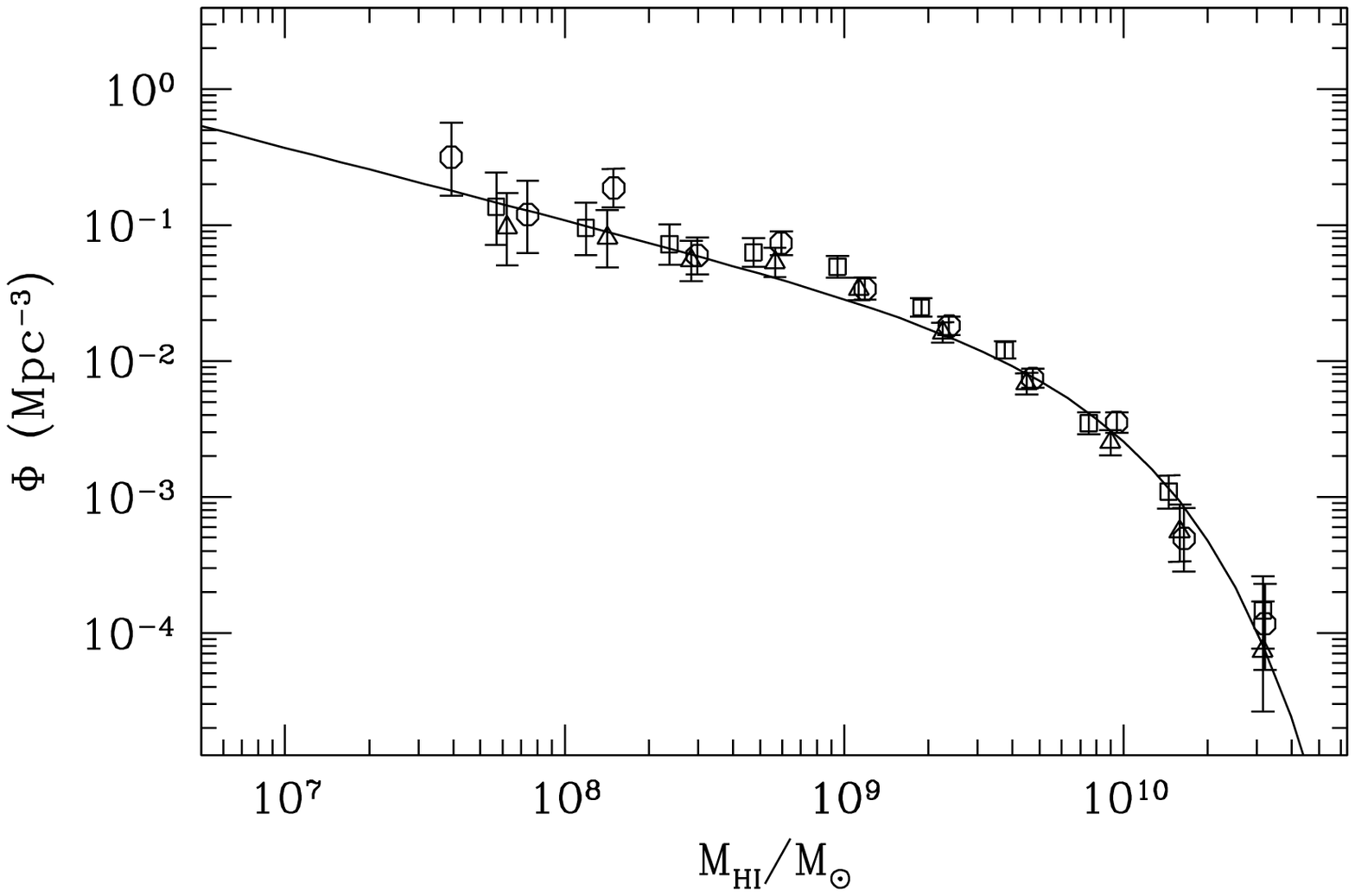}}}

\vspace{0.1in}
\figcaption{Effects on the H~I mass function of varying survey parameters.
The circles show the effect of using distances based on $V_0$
instead of the flow-corrected model of Tonry et al. (2000).
The triangles show the result of narrowing the declination offset to $2'$.
The squares show how the $1/{\cal V}_{tot}$ results change if
a correction is made for the large scale structure determined
from optical data in the survey region (\S 5.1).
The Schechter function curve shows our same fit
as in the previous figure to aid comparisons.}
\vspace{0.2in} 

In Fig. 4, we show how the mass function is affected when we
alter some of our basic assumptions. Circles in the figure show that only
small changes result if we base distances on $V_0$, which makes only a
simple correction for Local Group motion (de Vaucouleurs et al.~1977),
instead of the Tonry et al. (2000) flow model. The use of $V_0$ appears to
make the mass function slightly steeper, but not significantly so. This
suggests that distance uncertainties are unlikely to radically affect our
results.

Triangles in the figure show the effect of limiting our coverage to
galaxies with declination offsets smaller than $\Delta\delta<2'$ (triangles
in figure). Within these smaller offsets, our sensitivity corrections
($f(\Delta\delta)$; see \S 2) are less than a factor of 2.

Similarly minor changes to the H~I mass function resulted when: (1) sources
were detected in regions of single coverage or double coverage; (2) the
minimum velocity cutoff for our volume calculations was changed from 200
\kms\ to 100 or 300 \kms; or (3) sources were detected with the 21 or 22 cm
feed. We note that most of the low mass sources were detected with the 21
cm feed which had better sensitivity at small redshifts. This is another
difference from the Zwaan et al.~(1997) data, which was based mostly
on 22-cm-feed data.

\subsection{Simulation Tests and the Eddington Effect}

In order to test our methodology, we conducted detailed simulations of our
entire procedure. We carried out these simulations to: (1) ascertain
whether our completeness correction procedure had any unexpected effects;
(2) address the possibility of biases that might have caused the $1/{\cal
V}_{tot}$ to yield a different mass-function slope from an input
population; and (3) determine the probable ranges for $\Phi$ in our
low-mass bins where we detect a small number of sources.

Because H~I surveys do not yet probe the extragalactic population very
deeply, biases may be introduced due to small number statistics or from
uncertainties in the distances. Besides testing our procedures, we wanted
to investigate the possibility that distance uncertainties might bias the
low-mass end of our derived function. For example, Schechter (1976)
discusses a bias in the shape of the luminosity function due to high mass
sources at larger distances having redshifts that erroneously imply they
are nearby and low-mass. Alternatively, low-mass sources may appear to have
higher masses, so the overall effect on the mass function is not obvious.

Schechter showed that the ``Eddington correction'' for using redshift as a
distance indicator can be quite large for faint objects observed at small
redshifts where their distances are very uncertain. Even though we apply a
flow correction model and exclude the core of Virgo from our derivation
(which Schechter did not do) the residual uncertainty in distances will
still most heavily influence the mass determination for low-mass sources
since they are predominantly nearby.

We generated samples of galaxies that obeyed an input Schechter function,
gave them H~I linewidth properties that mimicked our detected galaxies,
allowed for random inclinations, and located the galaxies randomly within
our survey volume. Gaussian noise was added to each source and we simulated
the detection steps, including effects for beam offset and frequency
response, to determine which sources were detected. We also simulated all
of the subsequent steps, including remeasurement of the H~I flux (with
appropriate levels of noise inserted), calculation of ${\cal V}_{tot}$ for
each source using the completeness roll-off described earlier, and finally
calculated the mass function for each set of galaxies.

To explore the Eddington correction, the distances we used in our
mass-function calculation were based on the sources' redshifts, to which we
added a Gaussian dispersion of 300 km s$^{-1}$. For sources outside the
core of a cluster, this 1-$\sigma$ uncertainty of 4 Mpc should represent a
conservative estimate of our distance uncertainty. If the uncertainty were
any larger than this we would have expected to detect some blueshifted
sources outside of the Virgo core.

Each simulation required $\sim$60,000 galaxies (down to a mass of $10^{6.5}
M_\odot$), in order to ``detect'' $\sim$233 galaxies, to match our observed
sample size. We ran 1000 simulations each for an input Schechter function
with a slope $\alpha=-1.53$ and for $\alpha=-1.2$, and used our same
reduction programs to derive the mass function from the resulting set
of detected sources.

Figure 5 shows the median source density derived for each
half-decade mass interval and 1- and 2-$\sigma$ (68\% and 95\%) confidence
intervals from the 1000 simulations. The dots show the median recovered 
value from a population of sources assumed to follow a Schecter function 
with (a) $\alpha$ = -1.53 and (b) $\alpha$ = -1.2. The simulations show no 
net offset
from the input mass function shape; they also provide some guidance for the
size of errors that might be caused by distance uncertainties and the
likelihood of detection of low mass sources below $10^{7.5} M_\odot$. 

It might at first appear surprising that we find no indication of the
Eddington correction. The effect does become visible when we increase the
dispersion in redshifts to 600 \kms, a dispersion of 1000 \kms\ raised the
slope of the mass function from an input value of $\alpha=-1.2$ to a
derived value of about $-1.4$. These are much larger distance errors than
are plausible for our sample, so we conclude that the Eddington correction
is not significant for the ADBS.

The Eddington correction assumes, in effect, that more high-mass galaxies
will be scattered into low mass bins than vice versa. While there is a
larger volume from which to draw more massive sources that are Doppler
shifted to a particular observed velocity, there is a larger space density
of low-mass sources in the nearer volume. The result, at least for
plausible velocity dispersions, is that comparable numbers of sources are
scattered up and down into neighboring mass bins.

\centerline{\epsfxsize=0.9\hsize{\epsfbox{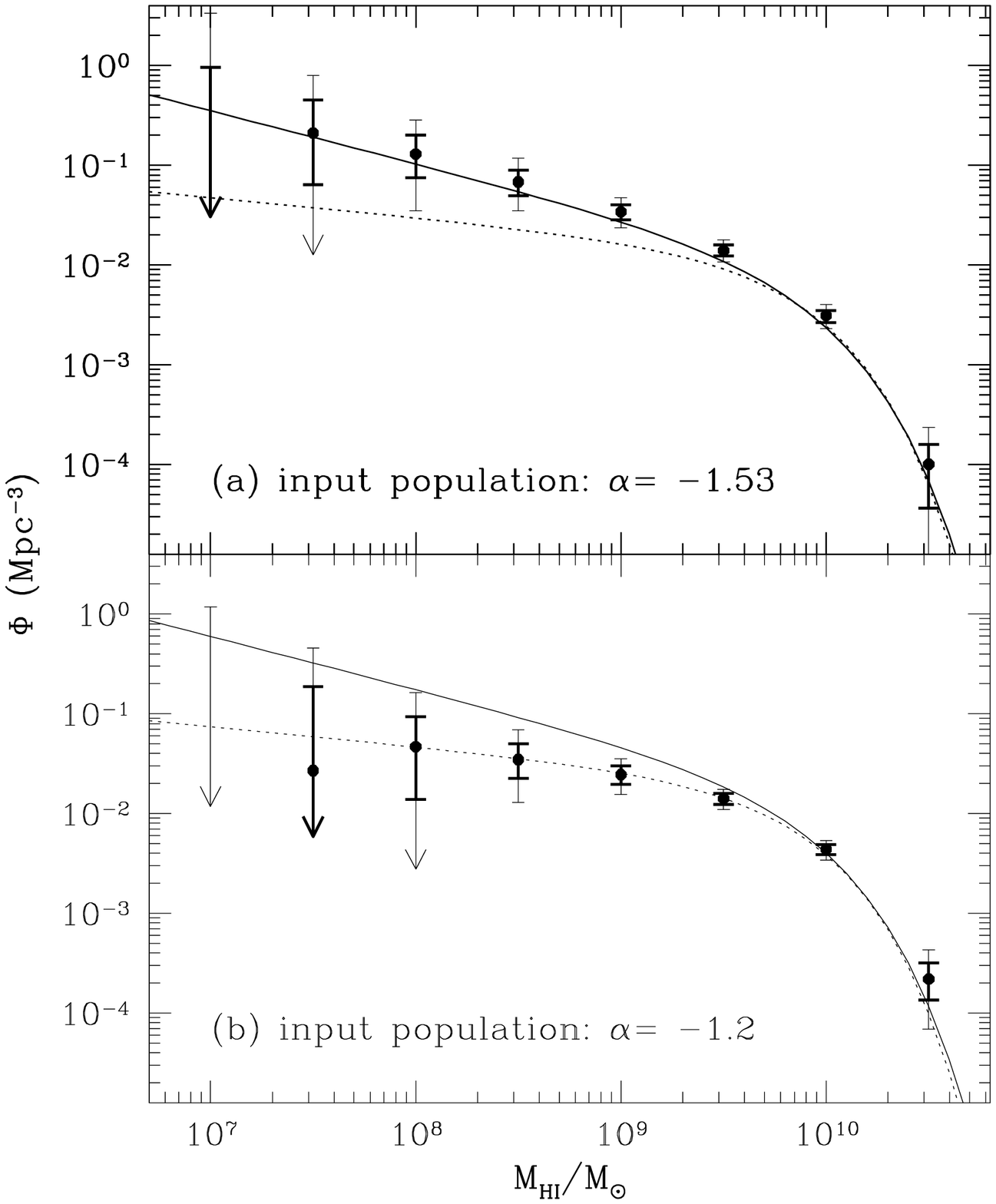}}}

\vspace{0.1in}
\figcaption{Statistical and distance-error uncertainties in the mass function.
The figure shows the range of mass function results recovered from 1000
simulations of the ADBS data. The dots show the median recovered value from a   
population of sources assumed to follow a Schechter function with (a)   
$\alpha=-1.53$ or (b) $\alpha=-1.2$, shown respectively by solid and dashed
curves in the figures. The simulations assumed a velocity
dispersion of 300 km s$^{-1}$, corresponding to distance uncertainties
of $\pm4$ Mpc. The heavy error bars show the $\pm1\sigma$ (68.3\%) range of 
results; the light
error bars show the $\pm2\sigma$ (95.4\%) range. For the lowest mass bins,
some of the expected values were zero, in which case these are marked as
upper limits.} 
\vspace{0.2in} 

Finally, we note that in the $\alpha=-1.2$ simulation, detection of $\sim$233
sources resulted in an overall density that was $\sim1.5\times$ higher than 
the Zwaan
et al.~(1997) mass function would have predicted for our search volume. In
the figure we have shifted the comparison mass functions (dashed and solid 
lines) upward so they pass
through the simulation points. This difference is expected because our mass
function results agree with Zwaan et al.~ at high masses, but we find a
larger number of low-mass sources per unit volume than their mass function
predicts. In other words, the Zwaan et al. mass function predicts that we
should have detected fewer sources within our search volume than we did.

\subsection{Effects of the Completeness Function on the Mass Function}

For the application of the SWML method to our data, we use only sources
with ${\cal S}_{obs}/{\cal N}_{eff}>7$, and we treat them as though they
would become undetectable at a distance where they would drop below this
level. Thus we make no use of our completeness roll-off results, which is
in keeping with the normal application of this method.

For the $1/{\cal V}_{tot}$ method, which has been used in all previous H~I
mass function studies,
we consider how the results might be affected by differences in the 
calculation of the total detection volume for each source.

In contrast with our empirical method of determining the completeness,
previous surveys have usually assumed that the effective noise depended on
linewidth as $w^{0.5}$, and have claimed a step-function for the
completeness behavior, often at ${\cal S}_{obs}/{\cal N}=5$. We discuss
here how such assumptions are likely to have changed the estimate of the
mass function relative to our analysis.

We believe that the most critical problem with earlier surveys is {\it not}
the details of the shape of the completeness function, but their basic
assessment of where their sensitivity ``cuts off.'' Some surveys have used
their faintest detected sources as an indication of a sensitivity limit.
Figure 1 shows that using the faintest detected sources to
determine this limit might imply completeness to a much lower level than
was actually achieved. The total potential detection volume ${\cal
V}_{tot}$ predicted for each source from such a low limit will then be
overestimated, and the mass function correspondingly underestimated.

The highest mass sources are bandpass limited, so exaggerating a survey's
sensitivity does not affect their predicted detection volume
much.\footnote{The volume estimates are still somewhat affected in a
driftscan survey like the ADBS or Zwaan et al.~(1997) because even
high-mass sources will be relatively weak at large declination offsets.}
However, low mass sources may only be detectable to slightly beyond the
minimum detection velocity, $v_{min}$, which is determined by such things
as confusion imposed by local Galactic high velocity clouds (assumed to be
200 \kms\ here). If the maximum detectable distance ($v_{max}/H_0$) is
overestimated, the predicted detection volume, ${\cal V}_{tot} \propto
(v_{max}^3-v_{min}^3)$ may increase more rapidly than even the cube of the
distance-overestimation factor. Thus, when the sensitivity of an H~I survey
is exaggerated, the assumed detection volume will be overestimated more for
low mass sources than high mass sources, causing the mass function to
appear too shallow.

Using the roll-off in completeness introduces only a small adjustment to
our estimate of the total volume in which a source could have been
detected. If we substitute a step-function at the point where we reach 50\%
completeness, the total search volume we estimate for each source is always
within $<$20\% of the value calculated from our completeness-function-based
method, and overall has a negligible effect on the mass function. Likewise,
using the linewidth dependence that we have established empirically only
makes a small change from assuming a $w^{0.5}$ noise dependence. In fact,
if we assume a $w^{0.5}$ noise dependence, the slope of our derived mass
function becomes marginally steeper.

\vfill
\section{Is the H~I Mass Function Affected By Large Scale Structure?}

The question labeling this section has two aspects: (1) Does large scale
structure introduce a bias into our estimate of the H~I mass function? (2)
Are there variations in the shape of the mass function in different density
regions? In \S 5.1 we show that large scale structure {\it does not} bias
our HI mass function results. In \S 5.2 we show that the shape of the HI
mass function, on the other hand, {\it does} vary in different density
regions.

\subsection{Density Corrections}

Concerns about the effect of large scale structure within our search volume
should be largely eliminated by the use of the SWML method. We have noted,
though, that there are potential problems with this method if the shape of
the mass function varies with the local density of galaxies.

To understand possible effects of large scale structure we have used
redshift data for optically cataloged galaxies to conduct a check on the
amount of density variation (as a function of redshift) within our main
survey region: $18.0<\delta<28.7^\circ$ at all right ascensions and
$8.0<\delta<15.7^\circ$ at $0<\alpha<16^h$ (see Paper 1). We used galaxies
with photographic magnitudes brighter than $m<14.5$ after correcting for
Galactic extinction (Schlegel et al. 1999), drawn from the NASA/IPAC
Extragalactic Database (NED). We examined the NED sources in narrow
redshift ranges, and found that the shape of the optical luminosity
function fit a Schechter function with: $\alpha = -1.05$, $M_\ast = -19.9$.
After correcting the velocities using the same method as for the H~I
measurements (Tonry et al. 2000), we derived the optical source density at
each redshift by comparing the actual number counts to the predicted counts
from the luminosity function. The resulting run of density with corrected
velocity is shown in Fig. 6.

\centerline{\epsfxsize=0.9\hsize{\epsfbox{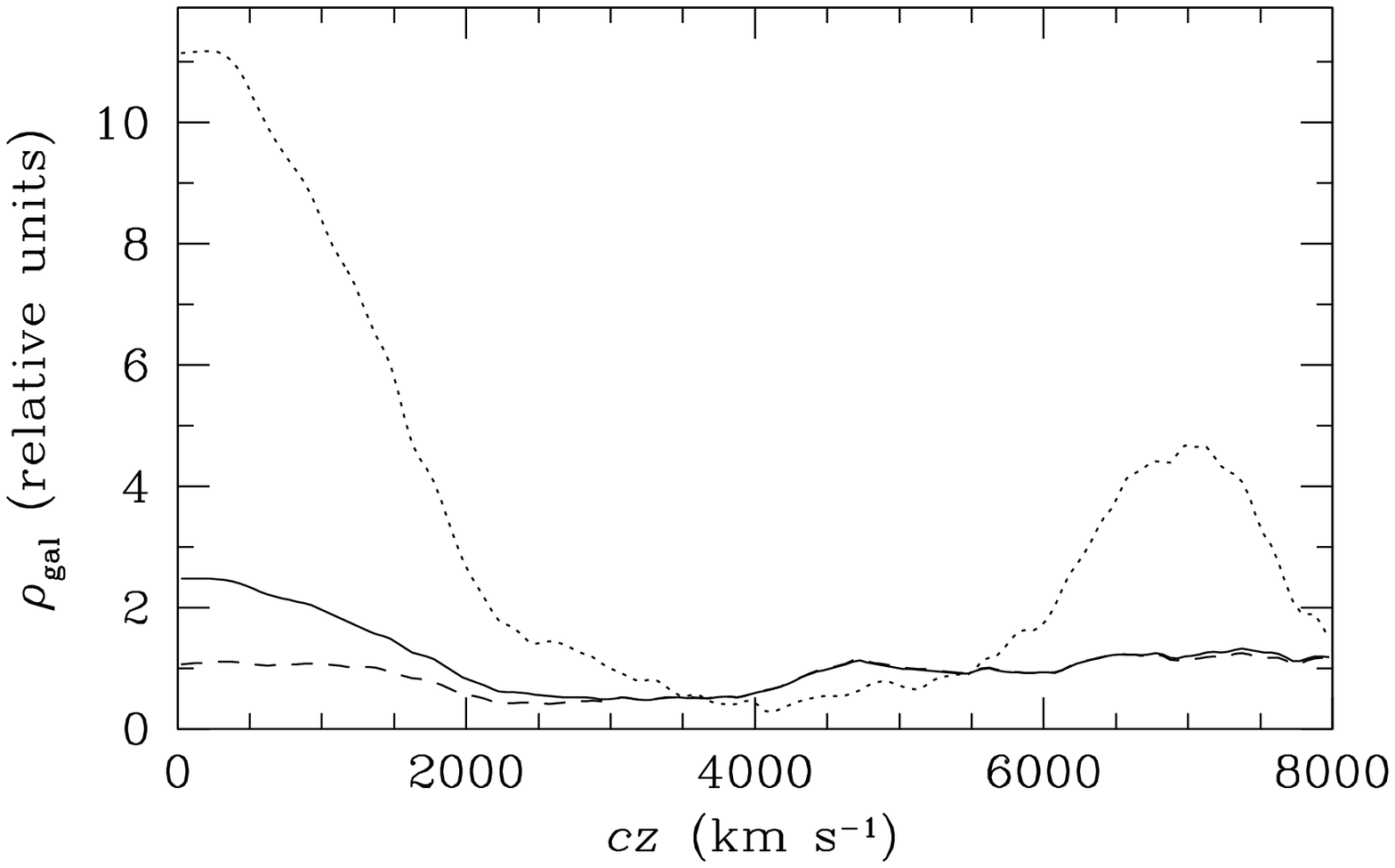}}}

\vspace{0.1in}
\figcaption{Density distribution of galaxies within the survey region based 
on optically selected galaxies. The solid line shows the mean density as a
function of redshift for the full ADBS. The dashed line shows the density
outside the core of Virgo. The dotted line shows the density within a
27$^\circ$ radius from the center of Virgo.}
\vspace{0.2in} 

Based on the optical sample of galaxies, we find that the density within
the ADBS survey region (excluding galaxies within $6^\circ$ of the core of
Virgo) is fairly uniform except within the interval $2000<cz<4000$ \kms\
where the density is about half of the other regions (solid line in
figure). The higher density at other redshifts reflects (a) the local
supercluster at small redshifts, (b) the Pisces-Perseus supercluster at
intermediate redshifts, and (c) the Coma cluster and associated ``great
wall'' region at high redshifts. It is important to note that our survey
region contains a mix of high and low density regions that average out to a
fairly uniform density.

In the slightly underpopulated interval of $2000<cz<4000$ \kms, the ADBS is
primarily sensitive to galaxies with H~I masses in the $8.5 <
\log(M_{HI}/M_\odot) < 9.5$ range. The effect of the lower density will
tend to suppress the counts in this portion of the mass function when using
the $1/{\cal V}_{tot}$ method. We can attempt to make a density correction
to the $1/{\cal V}_{tot}$ mass function using the optically-based density.
We modified the ${\cal V}_{tot}$ integral (\S4.1), multiplying it by the
density of sources at each velocity---essentially treating the higher
density regions as having a higher detection probability (see also SSR).

The effect of this correction is shown in Fig. 4 with square
symbols. Again the changes to the mass function are minor, although there
is a slight elevation of source counts in estimated density of sources
around $10^9 M_\odot$ as expected. After making these density corrections,
the overall normalization of the mass function rises slightly to
$\Phi_\ast = 0.0058$ Mpc$^{-3}$.

Using optical estimations of the density to correct the H~I-selected source
counts is crude at best, but it gives us some indication of how the mass
function shape and normalization might change. We expect that H~I-selected
sources will be less clustered than optically-selected galaxies, since
galaxies in higher-density regions may be depleted of gas, so the method is
likely to {\it over-}correct any density problems. The lack of significant
changes to the $1/{\cal V}_{tot}$ mass function resulting from this density
correction therefore indicates that the results are unlikely to be much
affected by large-scale structure within the search region.

We also note that the low-H~I mass galaxies in the ADBS are at similar
or higher velocities than the sources in the previous Arecibo H~I surveys,
and the density structure we find here is more uniform than in those
surveys (see SSR). Therefore our density estimate for the lowest mass bin
should be less subject to bias than those earlier surveys.

\subsection{Effects of Environment}

The Arecibo Dual-Beam Survey covers a large swath of sky that includes the
Virgo Cluster. Within the Virgo region, a signficant overdensity of
galaxies is evident out to about $27^\circ$ from the center of the cluster
(dotted line in Fig. 6). In our primary analysis we
included galaxies outside of the $6^\circ$ core but within $27^\circ$ of
the center of Virgo. Including the galaxies in the outer regions of Virgo
yielded a more uniform overall density, while still avoiding the large
distance ambiguities associated with the core region. This gave a mix of
higher and lower density regions nearby, much as cluster and field regions
also contribute to our results at higher redshifts. The region out to a
radius of $27^\circ$ from the center of Virgo has a mean overdensity of
about a factor of 10. We examine here some possible changes in the
character of the H~I mass function within such a high density region.

Our sample of Virgo galaxies is small---38 galaxies within $27^\circ$ of the
center of the Virgo cluster, including 13 galaxies within the inner
$6^\circ$. More distant clusters are not useful for probing the H~I mass
function in high density regions because they are too far away for the ADBS
to have detected low mass sources. We caution that small number statistics
make our conclusions uncertain, but there are indications that the mass
function is less steep in high density regions.

\centerline{\epsfxsize=0.9\hsize{\epsfbox{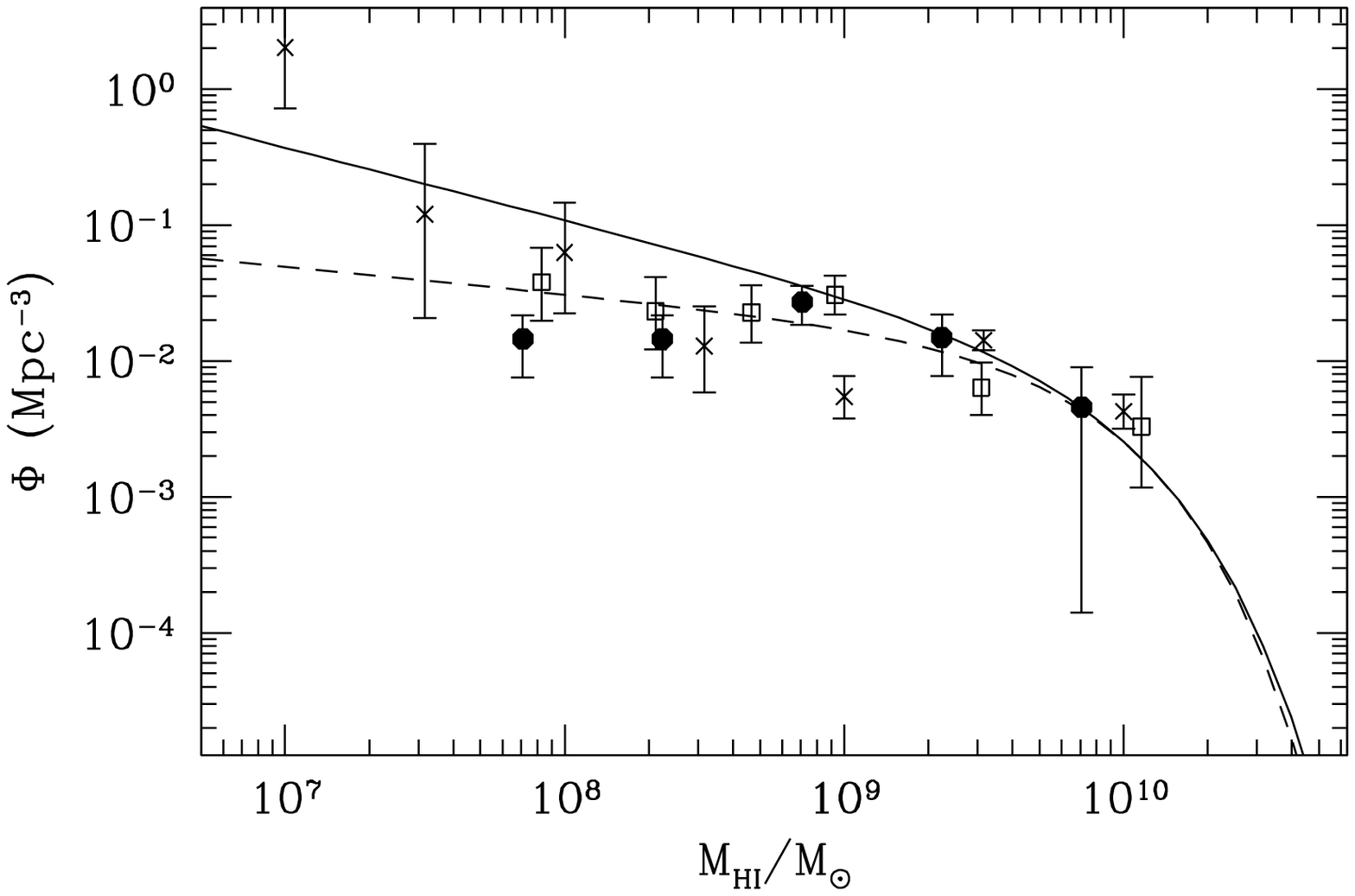}}}

\vspace{0.1in}
\figcaption{The H~I mass function within the Virgo cluster region.
The solid circles show the SWML estimate for galaxies within
$27^\circ$ of the center of the cluster and with redshifts
smaller than $cz<2300$ \kms. The open squares show the $1/{\cal V}_{tot}$
results over the same area. The solid and dashed curves are the same
as in Figure 2. The data from the Arecibo Slice survey
(Schneider et al. 1998) are shown as gray $\times$s. These data match the
shallower Schechter fit where the data are being drawn from a cluster
region (at the high mass end) and the steeper Schechter fit where they are
being drawn from a lower density region (at the low mass end).}
\vspace{0.2in} 

We calculate the mass function using the same methods as we described above.
Both the SWML and $1/{\cal V}_{tot}$ methods suggest a flatter mass
distribution. This result was found whether we assigned all of the galaxies
inside $6^\circ$ with redshifts smaller than $cz<2300$ \kms\ to a fixed
distance of 16.8 Mpc or used the solution from the Tonry et al. (2000) flow
model. In the figure, the SWML results use the flow model distances while
the $1/{\cal V}_{tot}$ results are based on a fixed distance for galaxies
in the core region. We applied the density correction described in \S5.1 to
the $1/{\cal V}_{tot}$ results. If we do not apply this correction, the
normalization of our mass function would make the density higher than our
earlier mass function at all masses, and about $10\times$ higher overall.
Accounting for the factor of 10 overdensity in the normalization allows us 
to more directly compare the results.

The differences in the mass function with galaxy density may help explain
the difference between our current mass function (solid line) and our
earlier estimate based on the Arecibo Slice of Spitzak \& Schneider (1998)
which is shown by gray $\times$ symbols in Figure 7. There
is a problem with the distribution of galaxies in the Arecibo Slice (see
Figure 8: all of the low mass and none of the high mass
galaxies are located at low redshifts while the opposite is true at high
redshifts. 

The relatively limited area covered in the Arecibo Slice was dominated by
the Pisces-Perseus supercluster at higher redshifts, so the mass function
for H~I masses greater than $\sim10^{8.5}$ is dominated by cluster
galaxies. H~I masses lower than $\sim10^{8.5}$ are drawn from lower density
regions. The ``turn up'' that appears to occur at low masses in the Arecibo
Slice is consistent with the higher mass points being drawn from a high
density region and the low mass points from a low density region. The
lowest mass points are consistent with a slope of $\alpha \approx -1.5$ while the
higher mass points, drawn from the higher density regions, are consistent
with a slope of $\alpha = -1.2$. The joining of these two nearly
independent H~I mass functions produces the apparent ``turn-up" in the
Arecibo Slice H~I mass function.

\centerline{\epsfxsize=0.9\hsize{\epsfbox{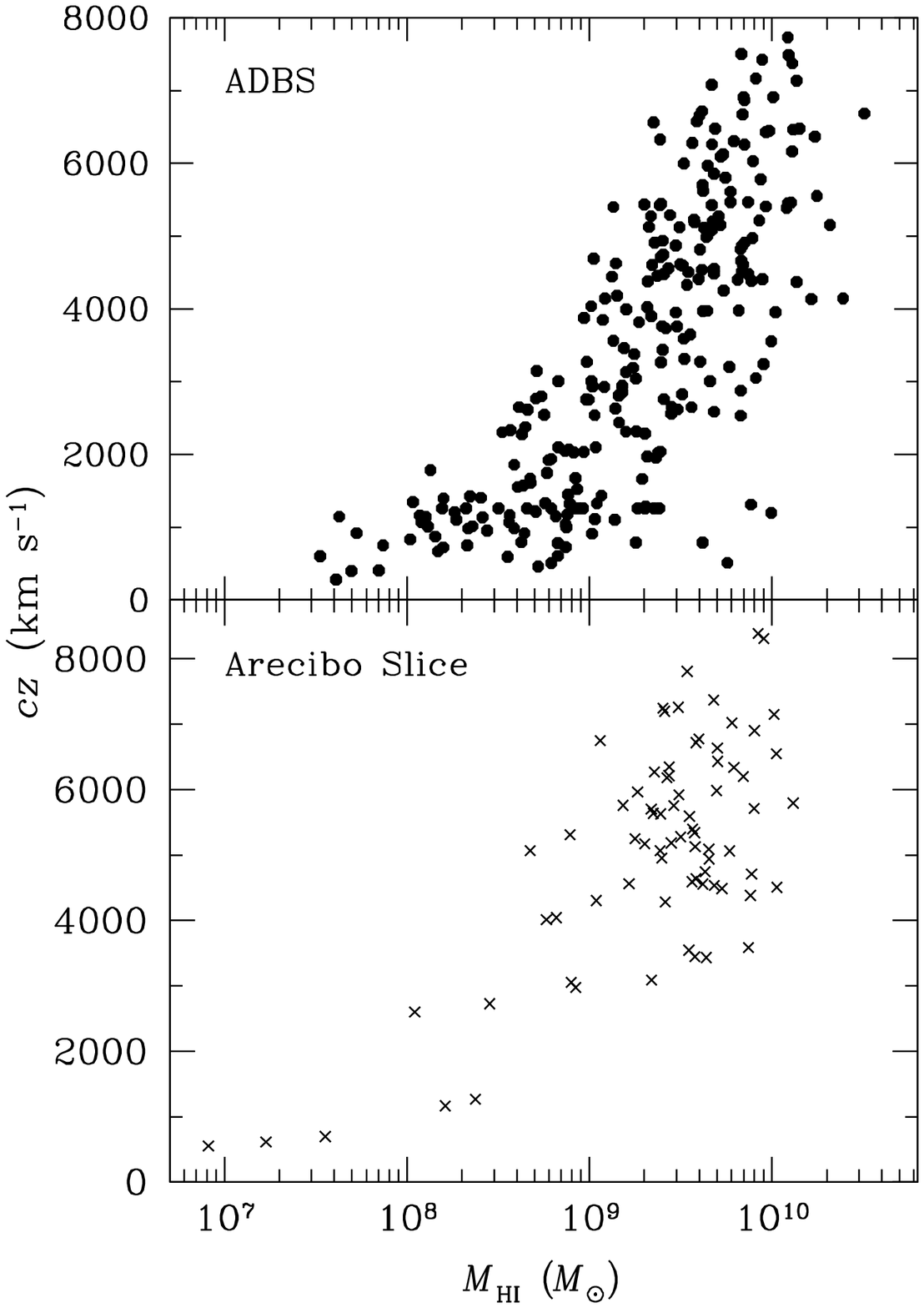}}}

\vspace{0.1in}
\figcaption{The redshifts of galaxies as a function of their H~I mass in 
(bottom panel)
the Arecibo Slice (Spitzak \& Schneider 1998) and (top panel) the ADBS.}
\vspace{0.2in} 

By contrast, Figure 8 demonstrates that the ADBS covered a
large enough area that several high mass galaxies were detected at lower
redshifts. The larger range of distances over which high mass galaxies are
detected helps to anchor the relative number density of high and low mass
galaxies.

\section{Limits on Ultra-High-Mass H~I Sources}

The galaxy Malin 1, identified by Bothun et al. (1987), has been pointed to
as evidence that we might be missing a significant population of high mass,
low surface brightness galaxies (Disney et al. 1987). Large, H~I-rich but
low surface brightness galaxies are extremely difficult to detect with
standard optical techniques. H~I surveys that cover large enough volumes
are ideal for identifying these galaxies since they are as easy to detect
as any other galaxy with a large H~I mass.

The ADBS data confirm that a large population of massive H~I-rich galaxies
is not lurking just below the sky surface brightness limits. While our
statistics at the high mass end of the mass function are poor, we are
sensitive to these galaxies in a large volume ($\sim8\times 10^5$ Mpc$^3$)
as seen by the fact that we detect high-mass sources with offsets as large
as $12'$ from the center of the main beam. The mean fluxes of these
galaxies are $\sim$10\% of their remeasured values, but with our
sensitivity limits, we should be able to detect galaxies in excess of
$5\times10^{10}$ M\solar\ at these large offsets out to the limiting
redshift of our survey.

With our ability to detect sources this far from the midline of the
drift-scans, the declination strips nearly overlap (the strip separations
alternate between 0.6$^\circ$ and 0.4$^\circ$), which corresponds to a
volume coverage of 8$\times 10^5$ Mpc$^3$ over an area of $\sim$1 sr for
these high mass systems. Nevertheless, we detect no high mass, low surface
brightness galaxies. We do not worry about column density constraints in
quoting this number because even a Malin-1-like galaxy would not be
resolved in most of the volume. For very nearby giants, we are only
sensitive down to a column density $> 2 \times 10^{19}$ cm$^{-2}$.

In the nearby regime, the Zwaan et al. (1997) survey places a much tighter
constraint on the population of high mass, low surface brightness galaxies
because it is sensitive to column densities $> 2 \times 10^{18}$ cm$^{-2}$.
These statistics restrict the population of Malin 1-like galaxies to
$<5.5\times10^{-6}$ Mpc$^{-3}$.

\section{Summary and Discussion}

We have used the ADBS to study the mean H~I mass over a wide range of
environments, and find a steep-sloped Schechter function: $\alpha = -1.53$,
$M_\ast = 9.88 M\solar$, $\Phi_\ast = 0.0058$ Mpc$^{-3}$ (the normalization
is lower, $\Phi_\ast = 0.0048$ Mpc$^{-3}$, when a density correction for
large scale structure effects is not applied). The ADBS mass
function results have been derived with very careful attention to the
sensitivity function. We inserted ``synthetic" sources which underwent all
of the data processing procedures to allow us to derive the sensitivity as
a function of H~I line width and to derive the completeness function.
Additionally, we find that our mass function results are robust to changes
in an assortment of parameters such as the minimum distance and velocity
flow models, and are consistent whether we use the $1/{\cal V}_{tot}$ or
SWML method.

Our mass function differs significantly from some previous determinations,
but this is probably because, in large part, previous H~I mass functions
were derived from optically selected samples or from samples that surveyed
high density regions. It appears likely that the mass function has a
different shape in clusters and in the field. There is evidence from
optical surveys that the luminosity function evolves with time (Lin et al.
1999, Sawicki et al. 1997) and density may also affect the shape of the
mass and luminosity functions (Phillipps et al. 1998; Wilson et al. 1997).
Gas stripping, evolution, and the merger rate of galaxies, which are more
significant in higher density environments, may preferentially remove gas
from low mass systems or destroy them altogether. We find that the mass
function in the Virgo cluster has a shallower faint-end slope,
$\alpha=-1.2$, similar to that found by Verheijen et al.~(2000) in the Ursa
Major region, which also has an overdensity of about a factor of 10.

It is interesting to note that the changes in $\alpha$ we find with density
may exhibit the opposite trend in optical samples. Phillipps et al. (1998)
find that the slope of the luminosity function gets steeper in the higher
density regions. This difference emphasizes that galaxies' gas mass and
optical luminosity are not directly related.

\centerline{\epsfxsize=0.9\hsize{\epsfbox{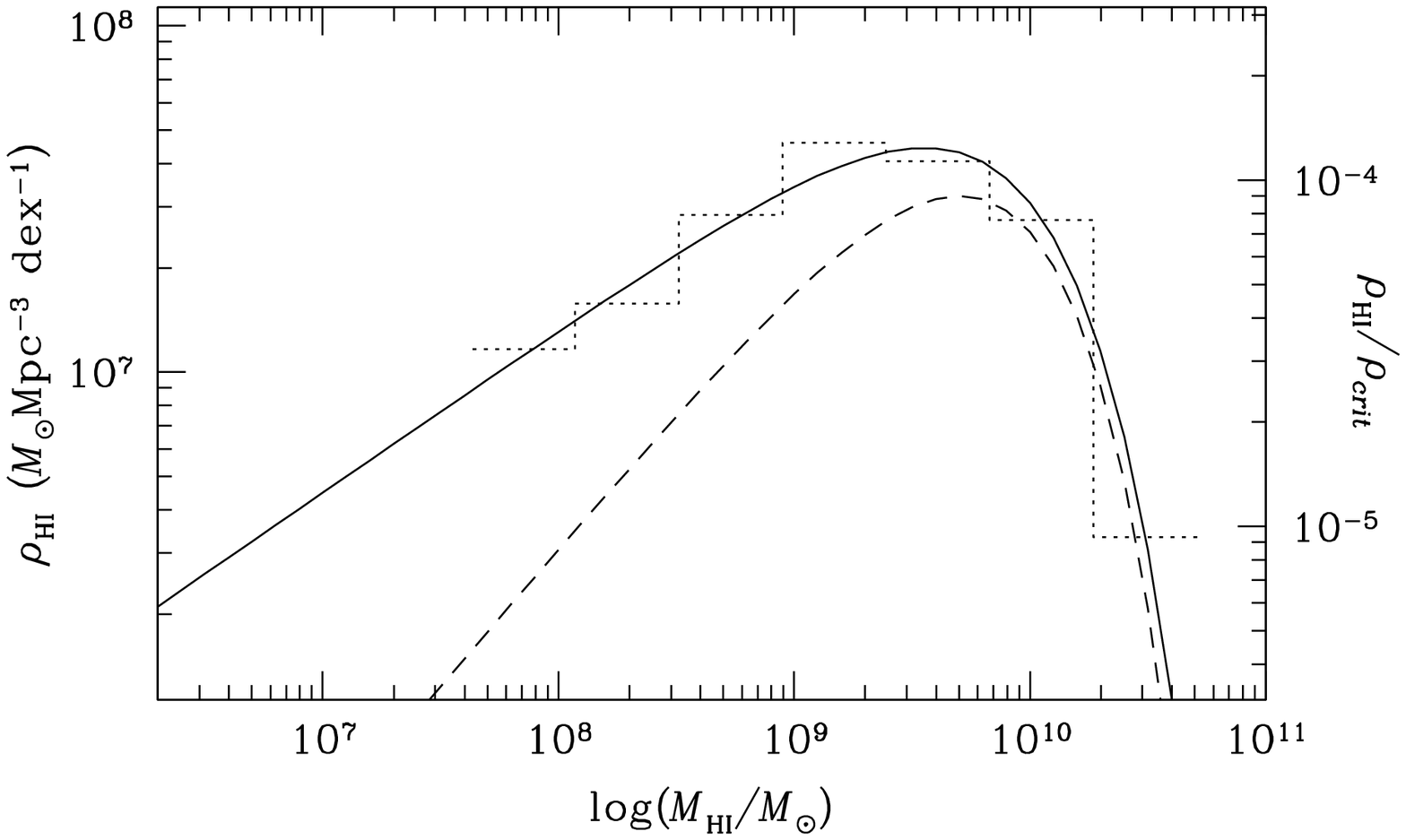}}}

\vspace{0.1in}
\figcaption{The relationship between H~I mass density (left-hand y-axis) or
$\Omega_{HI}$ (right-hand y-axis) and H~I mass. Most of the galaxy
contribution to the H~I mass density and $\Omega_{HI}$ is from $M_\ast$
galaxies, but the contribution from low mass sources is not negligible. The
solid line fit to the histogram shows our Schechter function, $\alpha$ =
-1.53. The dashed line shows the Zwaan et al. (1997) value, $\alpha$ = -1.2.
The integral of this curve represents $\sim$1\% of $\Omega_b$ that resides
in the H~I component of galaxies. }
\vspace{0.2in} 

Figure 9 shows the H~I mass density of galaxies ($\rho_{HI}$)
and the fraction of the critical density ($\rho_{HI}/ \rho_{crit}$)
contained in H~I-rich galaxies as a function of the galaxy's H~I mass. This
figure demonstrates that the mass contribution to ($\rho_{HI}/
\rho_{crit}$) is largest from galaxies near $M_\ast$. However, our
steep-sloped mass function indicates that the contribution from low mass
sources is not negligible. The value of $\Omega_b$ inferred from D/H
studies is 0.0445$h^2_{75}$ (Burles \& Tytler 1998). Given this value, we
find that $\sim$1\% (0.000484$h^2_{75}$) of the baryonic mass is contained
in the H~I within galaxies. By contrast, Penton et al. (2000) have shown
that $\sim$20\% of the baryons are tied up in low column density hydrogen,
primarily H~II, observed in the Lyman-$\alpha$ forest. The relative
contribution of the high and low density material suggests that H~I-rich
galaxies are relatively rare concentrations of neutral gas embedded in a
more substantial low density medium.

The ADBS is one of the largest surveys to date, particularly with respect
to low mass sources, yet the statistics at the low mass end remain thin.
Given practical limits for existing 21 cm telescopes, it seems likely that
this will remain a problem for some time to come. It will therefore be
vital for future surveys to carefully account for the completeness function
in their design. A particularly important question for future endeavors
will be the relationship between the shape of the mass function and galaxy
environment. To understand evolutionary processes in galaxies, we will have
to establish how the mass function changes with environment and with time.

\begin{acknowledgements}

The Digitized Sky Surveys were produced at the Space Telescope Science
Institute under U.S. Government grant NAG W-2166. The images of these
surveys are based on photographic data obtained using the Oschin Schmidt
Telescope on Palomar Mountain and the UK Schmidt Telescope. The plates were
processed into the present compressed digital form with the permission of
these institutions.

This research has made use of the NASA/IPAC Extragalactic Database (NED)
which is operated by the Jet Propulsion Laboratory, California Institute of
Technology, under contract with the National Aeronautics and Space
Administration.

\end{acknowledgements}

\end{document}